\documentclass[12pt,journal,draftclsnofoot,letterpaper,oneside,onecolumn]{IEEEtran}
\usepackage{amssymb}
\usepackage{amsmath}
\usepackage{amsfonts}
\usepackage{graphicx}
\usepackage{subfigure}
\usepackage{cite}
\usepackage{enumerate}

\begin{document}

\newtheorem{condition}{Condition}
\newtheorem{assumption}{Assumption}
\newtheorem{colloary}{Colloary}
\newtheorem{theorem}{\bf Theorem}
\newtheorem{property}{\bf Property}
\newtheorem{proposition}{Proposition}
\newtheorem{lemma}{\bf Lemma}
\newtheorem{example}{Example}
\newtheorem{notation}{Notation}
\newtheorem{definition}{\bf Definition}
\newtheorem{remark}{Remark}

%\title{\Large{Truthful Mechanisms for Secure Communication in Wireless Cooperative Networks}}
\title{\huge{Dynamic Popular Content Distribution in Vehicular Networks using Coalition Formation Games}}

\author{
\IEEEauthorblockN{\normalsize{Tianyu Wang}\IEEEauthorrefmark{1}, \normalsize{Lingyang Song}\IEEEauthorrefmark{1}, \normalsize{Zhu
Han}\IEEEauthorrefmark{2}, \normalsize{and Bingli Jiao}\IEEEauthorrefmark{1} \\}
\IEEEauthorblockA{\IEEEauthorrefmark{1}\normalsize{School of Electrical Engineering and Computer Science, Peking University,
Beijing, China.} \\
\IEEEauthorrefmark{2}\normalsize{Electrical and Computer Engineering Department, University of Houston, Houston, USA.} }
}

\maketitle

\thispagestyle{empty}

\begin{abstract}
%\boldmath

Driven by both safety concerns and commercial interests, vehicular ad hoc networks (VANETs) have recently received considerable attentions. In this paper, we address popular content distribution (PCD) in VANETs, in which one large popular file is downloaded from a stationary roadside unit (RSU), by a group of on-board units (OBUs) driving through an area of interest (AoI) along a highway. Due to high speeds of vehicles and deep fadings of vehicle-to-roadside (V2R) channels, some of the vehicles may not finish downloading the entire file but only possess several pieces of it. To successfully send a full copy to each OBU, we propose a cooperative approach based on the coalition formation games, in which OBUs exchange their possessed pieces by broadcasting to and receiving from their neighbors. Simulation results show that our proposed approach presents a considerable performance improvement relative to the non-cooperative approach, in which the OBUs broadcast randomly selected pieces to their neighbors as along as the spectrum is detected to be unoccupied.

\end{abstract}

\IEEEpeerreviewmaketitle

\newpage
%%%%%%%%%%%%%%%%%%%%%%%%%%%%%%%%%%
\section{Introduction}

\IEEEPARstart{V}{ehicular} ad hoc networks (VANETs) have been envisioned to provide increased convenience and efficiency to drivers, with numerous applications ranging from traffic safety, traffic efficiency to entertainment \cite{OW-2009,HL-2008}, especially after the advent of IEEE 802.11p and IEEE 1609 standards \cite{UM-2009}. One particular type of service, popular content distribution (PCD), has recently attracted lots of attentions, where multimedia contents are distributed from the roadside units (RSUs) to the on-board units (OBUs) driving through an area of interest (AoI) \cite{LYL-2011}. Examples of PCD may include: a local hotel periodically broadcasts multimedia advertisements to the vehicles entering the city on suburban highways; and a traffic authority delivers real-time traffic information ahead, or disseminates an update version of a local GPS map \cite{LYL-2011}. In brief, the proposed PCD is a local broadcasting service, in which the users are the vehicles (referred to as OBUs) passing through and the contents are multimedia files with large sizes.

In traditional cellular networks, downloading services, especially broadcasting services, are accomplished by direct transmissions from the base station to the mobile users. However, this scheme may be infeasible for PCD in VANETs, since the existence of stationary infrastructures (the RSUs) cannot be guaranteed, and even with infrastructures, the mobile users (the OBUs) may still fail downloading the entire contents due to high speeds of vehicles, deep fadings of wireless channels, and large sizes of popular contents (such as emergency videos \cite{PLOGL-2006}). Inspired by the peer-to-peer (P2P) protocols on the Internet \cite{Bittorrent,eDonkey}, which go beyond client-server protocols by letting a client also be a server, we introduce similar ideas for PCD in VANETs by allowing vehicle-to-vehicle (V2V) transmissions. Specifically, for the vehicles fail to download the entire contents directly from the RSUs, we propose a cooperative approach for them to construct a P2P network, in which popular content pieces can be efficiently exchanged among the OBUs. However, the well established P2P techniques on the Internet \cite{LCPSL-2005} should be carefully inspected for PCD in VANETs, given that:
\begin{enumerate}
    \item due to deep channel fading and co-channel interferences, wireless links in VANETs, relative to the wired links on the Internet, are very unreliable;
    \item due to unreliable links and high mobility of vehicles, network topologies of VANETs, relative to the static topology of the Internet, are ever-changing and highly unpredictable.
\end{enumerate}
Hence, the expected P2P protocols for PCD in VANETs are no longer the application layer protocols based on reliable transmissions. Instead, cross-layer protocols should be constructed, which jointly consider content requests, peer locations, channel capacities, potential interferences and adaptation to environmental changes.

In literature, there are many related works emerging recently. In \cite{OK-2004,BHMBM-2006}, the authors focused on applying IEEE 802.11 access points to inject data into vehicular networks, and introduced the connectivity challenges posed by such an environment. In \cite{XOW-2004}, the authors proposed an opportunistic dissemination scheme, in which the data will be exchanged whenever two vehicles are close enough for data transmission. However, this approach cannot avoid potential collisions in media access control (MAC) layer and may suffer from severe reduction of data rate. In \cite{NDPGS-2005}, the authors studied the cooperative schemes for downloading services in VANETs, in which they proposed SPAWN, a pull-based, peer-to-peer content downloading protocol that extends BitTorrent. However, the peer and content selection mechanisms have high overhead and are not scalable, especially when most of the vehicles are interested in downloading popular contents. In \cite{ZZJ-2009}, the authors proposed a cooperative medium access control (MAC) protocol, VC-MAC, for gateway downloading scenarios in vehicular networks. However, the considered ``broadcast throughput" is not content-aware but purely based on link quality, which may not reflect the network performance correctly. Recently, many researchers resort to network coding \cite{ACLY-2000,HMKKESL-2006} methods for content downloading services in vehicular networks. In \cite{LPYPG-2006}, the authors proposed Code Torrent, a pull-based content distribution scheme using network coding, in which vehicles transmit passively upon the downloading requests initiated by others. In \cite{PLOGL-2006}, the authors proposed a push-based content delivery scheme using packet level network coding for emergency related video streaming. In \cite{LLLG-2008}, the authors provided an in-depth analysis of implementation issues of network coding in vehicular networks by considering general resource constraints (e.g., CPU, disk, memory) besides bandwidth. In \cite{PGLYM-2006}, the authors present CodeCast, a network-coding-based ad hoc multicast protocol for multimedia applications with low-loss, low-latency constraints such as audio/video streaming. In \cite{LYL-2011}, the authors proposed a push-based protocol using symbol level network coding for PCD. All the schemes using network coding, no matter the pull-based or the push-based, have improved the network performance by simplifying P2P transmissions. Further, for avoiding severe data collisions in wireless scenarios, the MAC layer schemes for coordinating P2P transmissions have also been proposed \cite{LYL-2011,PLOGL-2006}.

In this paper, we address PCD in VANETs from a game theory point of view using the coalition formation games. The coalition formation game, in which the players form coalitions to improve their individual profits \cite{Myerson-1991}, has recently been used in vehicular networks \cite{HNSBH-2011}, e.g., for RSUs cooperation \cite{SHHNH-2011} in content downloading and for bandwidth sharing in vehicle-to-roadside (V2R) communications \cite{NWSH-2010}. We propose a coalition formation game model, in which the overall performance of the average delay has been formulated by a utility function that all players aim to maximize. Combining the content requests, peer locations, channel capacities, and potential interferences in a single utility function, we jointly consider the simplification issue and the coordination issue for P2P transmissions. With an algorithm that converges to a Nash-stable equilibrium proposed, we present our entire approach for PCD in VANETs. From the simulation results, we show our proposed approach achieves a considerable performance improvement relative to the non-cooperative approach, in which the OBUs broadcast randomly selected pieces to their neighbors as along as the spectrum is detected to be unoccupied.

The rest of this paper is organized as follows. Section \uppercase\expandafter{\romannumeral2} provides the system model. In Section \uppercase\expandafter{\romannumeral3}, we model PCD in VANETs as a coalition formation game with transferable utilities by defining a utility function that reflects the network performance in the average delay. In Section \uppercase\expandafter{\romannumeral4}, with some mathematical concepts introduced, we propose a coalition formation algorithm for the game, and then give the proposed approach for the entire problem. In Section \uppercase\expandafter{\romannumeral5}, simulation results in various conditions are presented. Finally, we conclude the paper in Section \uppercase\expandafter{\romannumeral6}.

%%%%%%%%%%%%%%%%%%%%%%%
\section{System Model}%
%%%%%%%%%%%%%%%%%%%%%%%

\begin{figure}[!t]
\centering
\includegraphics[width=4.2in]{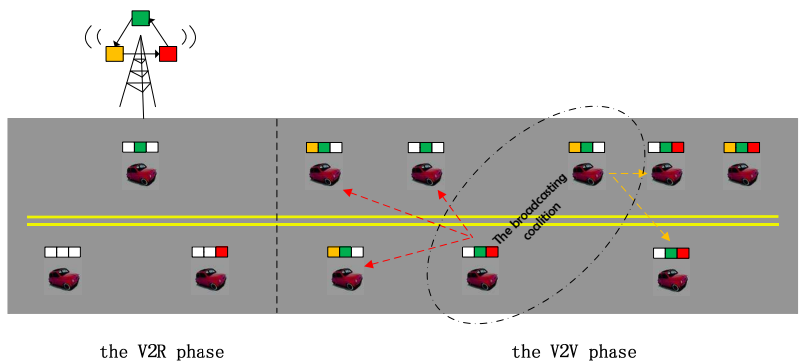}
\caption{System model of popular content distribution in vehicular ad hoc netowrks.} \label{system_model}
\end{figure}

Consider a vehicular ad hoc network consisting of $1$ RSU and $N$ OBUs, where the OBUs (the set of which is denoted by $\Omega$) are passing the RSU along a straight 2-lane highway, as shown in Fig.~\ref{system_model}. For any OBU, the entire process of PCD can be divided into two phases: the V2R phase and the V2V phase. In the V2R phase, the OBU is in the coverage of V2R transmissions, and keeps receiving popular pieces that are periodically broadcasted by the RSU. We assume the entire popular file is beyond the ability of V2R transmission in the V2R phase, which implies that any OBU can only obtain a fraction of the popular file after passing the RSU. In the V2V phase, the OBUs exchange the content pieces by broadcasting to and receiving from the others. In our scenario, no centralized channel coordinator exits and all OBUs compete for the same broadcasting channel, which is also used by the RSU for broadcasting in the V2R phase. The evolution of the popular file in each OBU is illustrated in Fig.~\ref{system_model}. Although the system model is quite simple, our proposed approach is independent from the traffic model and channel model, and thus, can be extended to more complicated scenarios. We will discuss the mobility model, channel model and content distribution model in the following subsections.

\subsection{Mobility Model}

The mobility model we use is similar to the Freeway Mobility Model (FMM) proposed in \cite{MPGW-2006}, which is well accepted for modeling the traffic in highway scenarios. In FFM, the simulation area includes many multiple lane freeways without intersections. At the beginning of the simulation, the vehicles are randomly placed in the lanes, and move at the history-based speeds. The vehicles randomly accelerate or decelerate with security distance $d_{min}>0$ maintained between two subsequent vehicles in the same lane, and no change of lanes is allowed.

In our scenario, the map has been simplified to a straight 2-lane highway as shown in Fig.~\ref{system_model}. All the OBUs independently choose to speed up or slow down by probability $p$ and acceleration $a>0$. The velocity of any OBU $i\in \Omega$ is limited by $v_{min} \le v_i(t) \le v_{max}$ for all time. To better reflect the changing topology of VANETs, we decide to allow the change of lanes when a vehicle is overtaking, as long as the security distance is maintained. Also, to reflect the car following issue, we give an upper bound $d_{max}$ for the distance between any two subsequent vehicles in the same lane. As our proposed model is for simplifying and coordinating P2P transmissions, the simulation begins at the V2V phase. The overall constraints are listed as follows:
\begin{enumerate}
    \item The OBUs are randomly placed on both lanes in an area with length $L$ and just leave the coverage of the RSU when the simulation begins.
    \item The initial speed of OBU $i \in \Omega$, denoted by $v_i(0)$, is randomly given in $[v_{min}, v_{max}]$.
    \item The speed of OBU $i \in \Omega$ satisfies:
            \begin{align} \label{Velocity}
            {v_i}(t+1) = \left\{
            \begin{array}{ll}
            {v_i}(t),& 1-2p, \\
            \min {\left({v_i}(t)+a,v_{max}\right)},& p, \\
            \max {\left({v_i}(t)-a,v_{min}\right)},& p, \\
            \end{array}
            \right.
            \end{align}
        where $0<p<1/2$ is the probability of acceleration and deceleration.
    \item For any OBU $i\in \Omega$ with OBU $j_1$ ahead in the same lane and OBU $j_2$ ahead in the other lane, OBU $i$ switches to the other lane, if ${d_{i,j_1}}(t) \le d_{min} $ and ${d_{i,j_2}}(t) > d_{min}$, or OBU $i$ decelerates to $v_i(t+1) = v_{min}$, if ${d_{i,j_k}}(t) \le d_{min}, k=1,2$.
    \item For any OBU $i\in \Omega$ with OBU $j_1$ ahead in the same lane and OBU $j_2$ ahead in the other lane, OBU $i$ accelerates to $v_i(t+1) = v_{max}$, if ${d_{i,j_1}}(t) \ge d_{max}$, or OBU $i$ switches to the other lane and accelerates to $v_i(t+1) = v_{max}$, if ${d_{i,j_1}}(t) < d_{max}$ and ${d_{i,j_2}}(t) \ge d_{max}$.
\end{enumerate}

\subsection{Channel Model}

For the vehicular channels, it is customary to distinguish between V2R and V2V channels. Generally speaking, these channels not only differ from each other, but also deviate significantly from those in cellular communication \cite{MMKTPBZKC-2011}. In the V2R phase, the antenna of the RSU is high enough that a line of sight (LOS) exits for any OBU $i\in \Omega$ in the coverage. We assume that the data rate of V2R transmission for any OBU $i\in \Omega$ is constantly $c_0$. In the V2V phase, the OBUs exchange data through V2V channels, which are highly affected by severe shadowing \cite{MMKTPBZKC-2011}. We assume the V2V link exists only between vehicles with a LOS, or equivalently between ``neighbors". All vehicles are equipped with a single antenna and the V2V transmission is divided into periodical slots with size $T$. In each slot, we adopt the Rician model for small-scale fading with the propagation loss factor $n=4$. The channel capacity between any two OBUs $i,j \in \Omega$ at slot $t$, is then given by
\begin{align} \label{V2VChannel}
c_{i,j}(t)=
\left\{
\begin{array}{ll}
W \log_2{\left(1+\eta \left|h\right|^2 d_{i,j}^{-4}(t)\right)}, & \mbox{LOS exits}, \\
0, & \mbox{otherwise}, \\
\end{array}
\right.
\end{align}
where $d_{i,j}$ is the distance between OBU $i$ and OBU $j$, $\eta$ is the signal-to-noise rate (SNR) at transmitters, $W$ is the channel bandwidth, and $h$ is the Rician channel gain given by
\begin{align} \label{Ricianchannel}
h = \sqrt{\frac{\kappa}{\kappa+1}} e^{j\theta} + \sqrt{\frac{1}{\kappa+1}} \omega,
\end{align}
where $\theta$ is a random variable uniformly distributed in $[0,2 \pi]$, $\omega$ is a complex Gaussian random variable with unit variance and zero mean, and $\kappa$ is the ratio of the energy in the LOS path to the energy in the scattered paths. Thus, the maximum data transmitted between OBU $i$ and OBU $j$ in slot $t$, is thus given by $c_{i,j}(t)T$.

\subsection{Content Distribution Model}

In the V2R phase, the RSU periodically broadcasts the popular file to the OBUs passing by. The popular file is equally divided into $M$ packets denoted by $\Gamma = \{\gamma_0,\ldots,\gamma_{M-1}\}$ with the size of each packet $s$. The OBUs keep receiving from the RSU when passing through with velocities $v_i(0), \forall i \in \Omega$. We assume the file size $Ms$ is beyond the maximum V2R throughput $c_0 D/v_i(0)$ for any OBU $i \in \Omega$, where $D$ is the diameter of the RSU coverage. Thus, the amount of initially possessed packets for any OBU $i \in \Omega$ is given by
\begin{align} \label{initialSegmentNumber}
n_i = \left[\frac{c_0 D}{v_i(0) s}\right].
\end{align}
As the popular packets are periodically broadcasted when the vehicles pass through the RSU, the indexes of the initially possessed packets for any OBU $i\in \Omega$ should be circularly continues, given by $\theta_i(\bmod{M}), (\theta_i+1)(\bmod{M}), \ldots, (\theta_i+n_i-1)(\bmod{M})$, where $0 \le \theta_i < M $ is the index of the first received packet. As the OBUs pass the RSU sequentially, for any OBU $j$ ahead of OBU $i$, we have
\begin{align} \label{firstReceivedIndex}
{\theta}_i = {\theta}_j +  \left[\frac{c_0 d_{i,j}(0)}{v_i(0) s}\right].
\end{align}

In the V2V phase, we assume that only one packet can be transmitted between any two OBUs in each slot, and the probability of successful transmission is proportional to the the ratio of channel capacity against the packet size, which is given by
\begin{align} \label{SuccessPossibility}
p_{i,j}(t) =
\left\{
\begin{array}{ll}
0, & T c_{i,j}(t) < s,  \\
\frac{T c_{i,j}(t)-s}{4s}, & s \le c_{i,j}(t) \le 5s,\\
1, & c_{i,j}(t)>5s, \\
\end{array} \right.
\end{align}
Also, considering data collisions in wireless network, OBU $i$ can not achieve any useful packet when more than one of its neighbors broadcast in the same slot.

%%%%%%%%%%%%%%%%%%%%%%%%%%%%%%%%%%%%%%%%%%%%%%%
\section{Coalition Formation Game}%
%%%%%%%%%%%%%%%%%%%%%%%%%%%%%%%%%%%%%%%%%%%%%%%

By focusing on the V2V phase, we consider the P2P transmissions in a given slot $t$. For any OBU $i \in \Omega$ in this slot, we denote the set of the possessed packets by $\Gamma_i(t)$, written as $\Gamma_i$ for short, and denote the set of ``neighbors" by $\Omega_i(t)$, written as $\Omega_i$ for short. Also, the probability of successful V2V transmission, denoted by $p_{i,j}(t)$ in (\ref{SuccessPossibility}), is written as $p_{i,j}$ for short. Next, we introduce a game theory model by defining a utility function that reflects the network performance as the average delay. For each OBU in the V2V phase, or equivalently each player in the game, it may choose to transmit or to receive in the current slot, only for maximizing its individual profit that is determined by the utility function.

\subsection{Utility Function}

\begin{figure}[!t]
\centering
\includegraphics[width=4.2in]{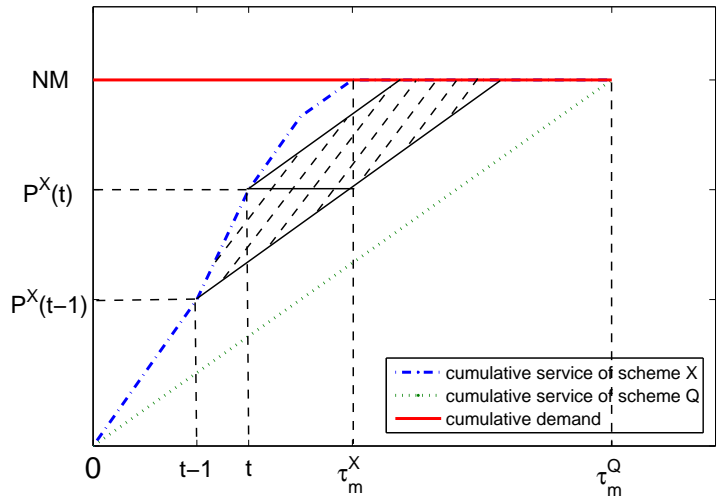}
\caption{The relative reduction in total delay of scheme $X$ compared with the standard scheme $Q$, where $NM$ is the constant demand curve, $P^X(t)$ is the service curve of scheme $X$ representing the packets already possessed at slot $t$, and $t_m^X, t_m^Q$ are the maximal delays for scheme $X$ and for the standard scheme $Q$, respectively.} \label{total_delay}
\end{figure}

In the proposed PCD problem, we consider the performance of the average delay experienced by the OBUs, which is generally defined as $\tau_a = \tau_t / N$ with $\tau_t$ representing the total delay experienced by all OBUs. For any given content distribution scheme $X$, $\tau_t^X$ is given by the area between the cumulative demand curve and cumulative service curve \cite{BG-1987}. In our scenario, we assume all arrivals of demand occur instantaneously at the beginning of the V2V phase and stay unchanged ever since \cite{QS-2004}, as shown in Fig.~\ref{total_delay}. In Fig.~\ref{total_delay}, we also show the cumulative service curves of two schemes $X$ and $Q$, and the details are given as below.

For a given content distribution scheme $X$, the average delay $\tau_a^X$ is given by
\begin{align} \label{averageDelay}
\tau_a^X = \frac{1}{N}  \sum\limits_{t=1}^{t^X_m} \left[NM - P^X(t) \right] T,
\end{align}
where $NM$ is the constant demand, $P^X(t) = \sum\nolimits_{i\in \Omega}{\left|\Gamma_i\right|}$ is the amount of total possessed packets in the current slot, and $t^X_m$ is the maximal delay defined by $P^X(t^X_m) = NM$ when each OBU in $\Omega$ has a full copy of the popular file.

For evaluating a given scheme, we define a standard scheme $Q$ with $P^Q(t) = t, 1 \le t \le NM$. In the standard scheme, only one packet is transmitted in each slot and the maximal delay is given by $t^Q_m = NM$. The average delay of $Q$ is then given by
\begin{align} \label{averageDelayStandard}
\tau_a^Q = \frac{1}{N}\sum\limits_{t=1}^{NM} \left( NM - t \right) T \approx \frac{NM^2T}{2}.
\end{align}

For any given scheme $X$ with the cumulative service function $P^X(t)$, it always enjoys reduction in the total delay relative to the standard scheme $Q$, if its service rate defined by $x(t) = P^X(t)-P^X(t-1)$ is greater than the service rate of the standard scheme $q(t) = P^Q(t)-P^Q(t-1) = 1$ for all slots. The corresponding reduction is given by the area between the cumulative service curve of scheme $X$ and the cumulative service curve of scheme $Q$, as shown in Fig.~\ref{total_delay}. In addition, this area can be divided into $t_m^X$ pieces with each piece representing the contribution of each slot. The piece representing the contribution of slot $t$ has been shaded in Fig.~\ref{total_delay}, the area of which can be calculated by dividing it into a triangle and a parallelogram sharing the same base $P^Q_{-1}[P^X(t)] - P^Q_{-1}[P^X(t-1)]$ but with different heights $x(t) = P^X(t)-P^X(t-1), NM-P^X(t)$, respectively. Here, $P^Q_{-1}(y) = y $ is the inverse function of $P^Q(t) = t$, and thus, the common base is derived as $P^X(t)-P^X(t-1) = x(t)$. Consequently, we have the contribution of slot $t$, the shaded area, is given by
\begin{align} \label{averageDelayReduction}
\Delta_{a}(t) = \frac{T}{N}\left\{ \frac{x(t)^2}{2} + \left[NM-P^X(t)-\frac{1}{2}\right]x(t) - [NM-P^X(t)]\right\},
\end{align}
where $x(t) = P^X(t)-P^X(t-1)$ is the service rate, representing the number of packets that are successfully delivered to the OBUs in slot $t$. Given a group of OBUs $S\subseteq\Omega$ representing the broadcasting vehicles in the current slot, we define the utility function $U(S)$ is proportional to $\Delta_{a}(t)$, which is thus given by
\begin{align} \label{utilityS}
U(S) = \alpha \left\{ \frac{x(t)^2}{2} + \left[NM-P^X(t)-\frac{1}{2}\right]x(t) - [NM-P^X(t)] \right\},
\end{align}
where $\alpha$ is a pricing factor. It can be proved that $U(S)$ is a positive increasing concave function with the independent variable $1 \le x(t) < NM-P^X(t)$.

To complete the expression of the utility function, the exact $x(t)$ is needed. Here, we propose a greedy algorithm for each OBU in $S$ to decide which packet to broadcast in the current slot. To be specific, any OBU $i \in S$ will broadcast $\gamma_k \in \Gamma_i$, only if the throughput for broadcasting $\gamma_k$ is statistically the largest for all possessed packets. We denote by $\Omega_i^* \subseteq \Omega_i$ as the set of OBU $i$'s neighbors that are not interfered by other transmitters and can therefore achieve useful data from OBU $i$, which is given by
\begin{align} \label{aliveNeighbors}
\Omega_i^* = \{ j\in\Omega_i | \left(\Omega_j \backslash \{i\} \right) \cap S = \emptyset \}.
\end{align}
The set of OBUs in $\Omega_i^*$ requesting for $\gamma_l$, denoted by $\Omega_i^*(\gamma_l)$, is given by
\begin{align} \label{aliveNeighborsWithoutASegment}
\Omega_i^*(\gamma_l) = \{ j \in \Omega_i^* | \gamma_l \notin \Gamma_j\}.
\end{align}
Thus, packet $\gamma_k \in \Gamma_i$ is broadcasted by OBU $i \in S$, if and only if
\begin{align} \label{segmentChosenCondition}
\sum\limits_{j \in \Omega_i^*(\gamma_k)} {p_{i,j}} \ge \sum\limits_{j \in \Omega_i^*(\gamma_l)} {p_{i,j}}, \forall \gamma_l \in \Gamma_i.
\end{align}
We denote by $\gamma_*^i$ as the packet resulting from the greedy algorithm for any OBU $i \in S$, Therefore, the expression of $x(t)$ is given by
\begin{align} \label{serviceRateExpression}
x(t) = \sum\limits_{i\in S}{\sum\limits_{j \in \Omega_i^*(\gamma_*^i)} {p_{i,j}}}.
\end{align}
The utility function (\ref{utilityS}) is then completed, and we can construct the game in the next subsection.

\subsection{Coalitional Game}

The P2P transmissions in the V2V phase are first modeled in a coalitional game with the transferable utilities \cite{Myerson-1991}, where the OBUs, as the game players, tend to form coalitions so that their individual profits are maximized. In other word, we have the following definition.

\begin{definition}\label{defcoalitionalgame}
A coalitional game with a \emph{transferable utility} is defined by a pair $(\Omega, V)$, where $\Omega$ is the set of players and $V$ is a function over the real line such that for every coalition $S \subseteq \Omega$, $V(S)$ is a real number describing the amount of value that coalition $S$ receives, which can be distributed in any arbitrary manner among the members of $S$.
\end{definition}

It is natural to treat the utility function in (\ref{utilityS}) as the value function $V(S)$. To achieve the value formulated in (\ref{utilityS}), the OBUs belonging to coalition S need to synchronize their communication and certain information needs to be collected from their neighbors, determining the broadcasting packets in the proposed greedy algorithm. Here, we consider a cost function that varies linearly with the size of the coalition $\left|S\right|$ as follows
\begin{align} \label{costFunction}
C(S)=
\left\{ \begin{gathered}
\beta \left|S\right|, \mbox{if}\left|S\right|>1, \hfill \\
0~~~~~, \mbox{otherwise}, \hfill \\
\end{gathered} \right.
\end{align}
where $\beta>0$ is a pricing factor. The motivation behind the cost function (\ref{costFunction}) is that, in order to synchronize to the network and determine the best packet for broadcasting, each of the OBU in coalition $S$ brings a constant cost (as the amount of neighbors for each OBU is stable), and thus, the entire coalition brings a cost proportional to the coalition size $\left|S\right|$.

Consequently, the value function of any coalition $S\subseteq \Omega$ is given by
\begin{align} \label{valueFunction}
V(S) = U(S) - C(S).
\end{align}
This function quantifies the total value that coalition $S$ receives, which should be distributed among all members of $S$ according to their individual contributions. Thus, the individual profit of any OBU $i \in S$ is given by
\begin{align} \label{payoffFunction}
\phi_i(S) = \frac{\sum\nolimits_{j \in \Omega_i^*(\gamma_*^i)} {p_{i,j}}}{\sum\nolimits_{j\in S}{\sum\nolimits_{j \in \Omega_i^*(\gamma_*^i)} {p_{i,j}}}} V(S).
\end{align}

In the proposed coalitional game, if all OBUs form a grand coalition $\Omega$ for broadcasting at the current slot, no vehicle can receive any useful data due to the sever interference. Thus, the service rate in the current slot is $x(t) = 0$. By substituting $x(t) = 0, |S| = N$ to (\ref{utilityS}), (\ref{costFunction}) and (\ref{valueFunction}), we have a negative value $V(\Omega) = -[NM-P(t)] -\beta N$ and then negative individual profits for any OBU in $\Omega$. Therefore, there is no motivation for the grand coalition $\Omega$ to form. Actually, the OBUs will deviate from the grand coalition and form independent disjoint coalitions. Hence, the proposed coalitional game is a $(\Omega, V)$ \emph{coalition formation game} \cite{SHDHT-2009}. In the next section, we will devise a coalition formation algorithm to achieve these disjoint coalitions.

%%%%%%%%%%%%%%%%%%%%%%%%%%%%%%%%%%%%%%%%%%%%%%%
\section{Coalition Formation Algorithm}%
%%%%%%%%%%%%%%%%%%%%%%%%%%%%%%%%%%%%%%%%%%%%%%%

In this section, we devise a coalition formation algorithm for the proposed coalition formation game, and then, propose the entire approach for PCD in VANETs. Also, we show that the proposed coalition formation algorithm converges to a Nash-stable partition, and the entire approach can adapt to environmental changes.

\subsection{Coalition Formation Concepts}

First, we introduce some necessary concepts, taken from \cite{BJ-2002}.

\begin{definition}\label{defcoalitionalstructure}
A \emph{coalitional structure} or a \emph{coalition partition} is defined as the set $\Pi = \{S_1,\ldots,S_l\}$, which partitions players set $\Omega$, i.e., $\forall k, S_k\subseteq \Omega$ are disjoint coalitions such that $\bigcup\nolimits_{k=1}^{l}\{S_k\} = \Omega$. Further, we denote by $S_{\Pi}(i)$, the coalition $S_k \in \Pi$, such that $i \in S_k$.
\end{definition}

One key approach in coalition formation is to enable the players to join or leave a coalition based on well-defined preferences. To be specific, each player must be able to compare and order its potential coalitions based on which coalition this player prefers to being a member of. For evaluating these preferences, the concept of preference relation or order is introduced.

\begin{definition}\label{defpreferencerelation}
For any player $i \in \Omega$, a \emph{preference relation} or \emph{order} ${\succeq}_i$ is defined as a complete, reflexive, and transitive binary relation over the set of all coalitions that player $i$ can possibly form, i.e., the set $\{S_k \subseteq  \Omega : i \in S_k\}$.
\end{definition}

Hence, for any given player, $i \in \Omega,S_1 {\succeq}_i S_2$ implies that player $i$ prefers being a member of a coalition $S_1 \subseteq \Omega $ with $i \in S_1$ over being a member of a coalition $S_2 \subseteq \Omega $ with $i \in S_2$, or at least, OBU $i$ prefers both coalitions equally. The preferences of the players could be quantified differently in different applications. In this paper, for any OBU $i\in \Omega$ and $ i \in S_1,S_2$, we propose the following preference
\begin{align} \label{preferenceRelation1}
S_1 {\succeq}_i S_2 \Leftrightarrow \phi_i(S_1) \ge \phi_i(S_2) ~\&~ {\phi}_j(S_k) \ge {\phi}_j(S_k \backslash \{i\}), \forall j \in S_k\backslash\{i\}, k =1,2.
\end{align}
This definition implies OBU $i$ prefers being a member of $S_1$ over $S_2$ only when OBU $i$ gains an increase in the individual profit and meanwhile no other OBUs in $S_1$ or $S_2$ suffers a decrease because of OBU $i$'s joining. The asymmetric counterpart of ${\succeq}_i$, denoted by ${\succ}_i$, is defined as
\begin{align} \label{preferenceRelation2}
S_1 {\succ}_i S_2 \Leftrightarrow \phi_i(S_1) > \phi_i(S_2) ~\&~ {\phi}_j(S_k) \ge {\phi}_j(S_k \backslash \{i\}), \forall j \in S_k\backslash\{i\}, k =1,2.
\end{align}
For any $S_k$ where $\exists j \in S_k\backslash\{i\}, {\phi}_j(S_k) < {\phi}_j(S_k \backslash \{i\})$, we say $S_k$ is the least preferred coalition. And we have $S_l{\succ}_i S_k$, if $S_l$ is not the least preferred, i.e., ${\phi}_j(S_l) \ge {\phi}_j(S_l \backslash \{i\}), \forall j \in S_l\backslash\{i\}$.

For forming coalitions from a given partition $\Pi$, we define the switch operation as follows.

\begin{definition}\label{defswitch}
Given a partition $\Pi = \{S_1,\ldots,S_l\}$ of the OBUs set $\Omega$, if OBU $i \in \Omega$ performs a switch operation from $S_{\Pi}(i) = S_m$ to $S_k \in {\Pi} \cup \{\emptyset\}, S_k \ne S_{\Pi}(i)$, then the current partition $\Pi$ of $\Omega$ is modified into a new partition ${\Pi}'$ such that ${\Pi}' = ({\Pi} \backslash \{S_m,S_k\}) \cup \{S_m \backslash \{i\}, S_k \cup \{i\} \}$.
\end{definition}

Finally, we define the history collection of OBU $i$ as follows.

\begin{definition}\label{defhistory}
Given any player $i \in \Omega$, the history collection $H(i)$ is defined as the set of coalitions that OBU $i$ visited and then left in the past.
\end{definition}

\subsection{Coalition Formation Algorithm}

We propose a coalition formation algorithm in which the OBUs form disjoint coalitions by switching operations. Specifically, given a partition $\Pi = \{S_1,\ldots,S_l\}$ of OBUs set $\Omega$, a switch operation from $S_{\Pi}(i) = S_m, m \in \{1,2,\ldots,l\}$ to $S_k \in {\Pi} \cup \{\emptyset\}, S_k \ne S_{\Pi}(i)$ is allowed for any OBU $i \in \Omega$, if and only if $S_k \cup \{i\} {\succ}_i S_{\Pi}(i)$ and $S_k \cup \{i\} \notin H(i)$. In this mechanism, every OBU $i \in \Omega$ can leave its current coalition $S_{\Pi}(i)$, and join another coalition $S_k \in {\Pi}$, given that the new coalition $S_k \cup \{i\}$ is strictly preferred over $S_{Pi}(i)$ through the preference relation defined in (\ref{preferenceRelation2}). The coalition formation game is summarized in Table \ref{coalitionFormationAlgorithm}, where the OBUs make switch operation in a random order. The convergence of this algorithm is guaranteed as follows.

\begin{proposition} \label{pro1}
Starting from any initial coalitional structure ${\Pi}_{initial}$, the proposed coalition formation algorithm maps to a sequence of switch operations, which will always converge to a final network partition ${\Pi}_{final}$ composed of a number of disjoint coalitions.
\end{proposition}

\begin{IEEEproof}
By carefully inspecting the preference defined in (\ref{preferenceRelation2}), we find that a single switch operation will either yields an unvisited partition, or a visited partition where one coalition degenerates to a singleton. When it comes to partition $\Pi$ where OBU $i$ forms a singleton, this non-cooperative OBU $i$ must either join a new coalition or decide to remain non-cooperative. If OBU $i$ decides to remain non-cooperative, then the current partition $\Pi$ cannot be changed to any visited partitions in the next round. If OBU $i$ decides to join a new coalition, then the switch operation made by OBU $i$ will form an unvisited partition without non-cooperative OBUs. In either case, an unvisited partition will form. As the number of partitions for a given set is the Bell number \cite{Ray-2007}, the sequence of switch operations will always terminate and converge to a final partition $\Pi_{final}$ after finite turns, which completes the proof.
\end{IEEEproof}

We study the stability of $\Pi_{final}$ by using the following concept from the hedonic games \cite{BJ-2002}.

\begin{definition}\label{defNash}
A partition $\Pi$ is Nash-stable, if $\forall i \in \Omega, S_{\Pi}(i) \succeq_i S_k \cup \{i\}$ for all $S_k \in \Pi \cup \{\emptyset\}$.
\end{definition}

\begin{proposition} \label{pro1}
The partition $\Pi_{final}$ in our coalition formation algorithm is Nash-stable.
\end{proposition}

\begin{IEEEproof}
Suppose the final partition $\Pi_{final}$ resulting from the proposed algorithm is not Nash-stable. Consequently, there exists an OBU $i\in \Omega$, and a coalition $S_k \in \Pi_{final}$ such that $S_k \cup \{i\} \succ_i S_{\Pi_{final}}(i)$. Based on our algorithm, OBU $i$ can perform a switch operation from $S_{\Pi_{final}}(i)$ to $S_k$, which contradicts the fact that $\Pi_{final}$ is the final partition. Thus, we have proved that any final partition $\Pi_{final}$ resulting from the proposed algorithm must be Nash-stable.
\end{IEEEproof}

\subsection{Popular Content Distribution Protocol}

The proposed coalition formation algorithm, as shown in Table \ref{coalitionFormationAlgorithm}, needs certain information to be transmitted among all the members in the network. Although the overhead is comparatively small to the data packets transmitted between vehicles, it still limits the network scale as the dynamic algorithm has to catch up the environmental changes, especially considering the potential splits in vehicular networks. Here, we propose a splitting scheme in Table \ref{split}, in which the OBUs automatically split into subnetworks, if the network scale surpasses a certain threshold $N_{max}$ or the OBUs split into disconnected parts.

Combining the splitting scheme in Table \ref{split}, the coalition formation algorithm in Table \ref{coalitionFormationAlgorithm}, and the greedy algorithm in Section \uppercase\expandafter{\romannumeral3}, we propose the entire approach for PCD in VANETs in Table \ref{contentDistributionProtocol}. In the V2R phase, the OBUs keep receiving packets from the RSU. In the V2V phase, the OBUs may first split into subnetworks due to scaling limits or location limits. Then, in each subnetwork, the coalition formation algorithm are performed by the OBUs to achieve efficient coalitions in the final partition. At last, the members of the most efficient coalition in each subnetwork, broadcast the packets resulting from the greedy algorithm in the current slot. By the periodical splits and calculations, the proposed approach adapts to the environmental changes in VANETs, and at the same time, achieves high performances in the average delay.

%%%%%%%%%%%%%%%%%%%%%%%%%%%%%
\section{Simulation Results}%
%%%%%%%%%%%%%%%%%%%%%%%%%%%%%

In this section, the performance of the proposed approach in Table \ref{contentDistributionProtocol} is simulated in various conditions compared with a non-cooperative approach, in which each OBU in $\Omega$ broadcasts a randomly chosen packet in each slot, as long as the channel is detected to be unoccupied. The simulation parameters are taken from a general highway scenario, as shown in Table \ref{parameters}.

\begin{figure}[!t]
\centering
\includegraphics[width=4.2in]{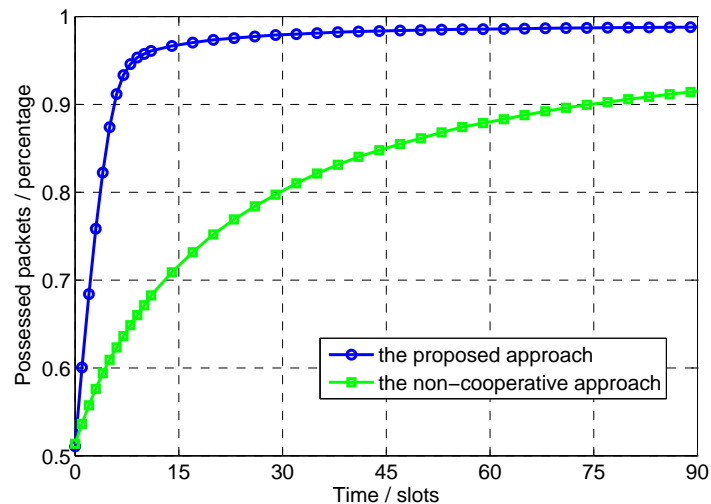}
\caption{Total possessed packets by the proposed approach and the non-cooperative approach as a function of time for networks with $N=8, L=800m, D=250m$.} \label{service_rate}
\end{figure}

In Fig.~\ref{service_rate}, we show the cumulative service curves of both the proposed approach and the non-cooperative approach for networks with $N=8, L=800m,$ and $D=250m$, where the vertical coordinate has been normalized by the total demand $NM$. First, we can see both the approaches have increasing service curves, while, the proposed approach performs much better than the non-cooperative approach. In the non-cooperative approach, each OBU makes individual decisions on whether to broadcast and what to broadcast, which may lead to inefficient broadcastings and severe data collisions. However, in the proposed approach, the OBUs cooperate with each other to maximize the utility function given in (\ref{utilityS}), which is highly dependent on the network throughput of the current slot. Consequently, the proposed approach achieves a better performance in service rate. Second, we find that the service rate given by (\ref{service_rate}) decreases with time for both approaches. At the beginning of the V2V phase, the initial possessed packets vary from vehicle to vehicle, and thus, the P2P transmissions are highly efficient. However, as the possessed packets tend to be the same for each OBU, the P2P transmissions become less efficient. Therefore, the service rate decreases with time for both approaches. Third, although the proposed approach has a high service rate in the first few slots, its cumulative service curve does not converge to $1$ even after $90$ slots. Actually, this can be explained by the splitting in vehicular networks, that some vehicle may depart from the rest before receiving the entire contents. Thus, the total possessed packets may not converge to $1$ even after arbitrarily long time.

\begin{figure}[!t]
\centering
\includegraphics[width=4.2in]{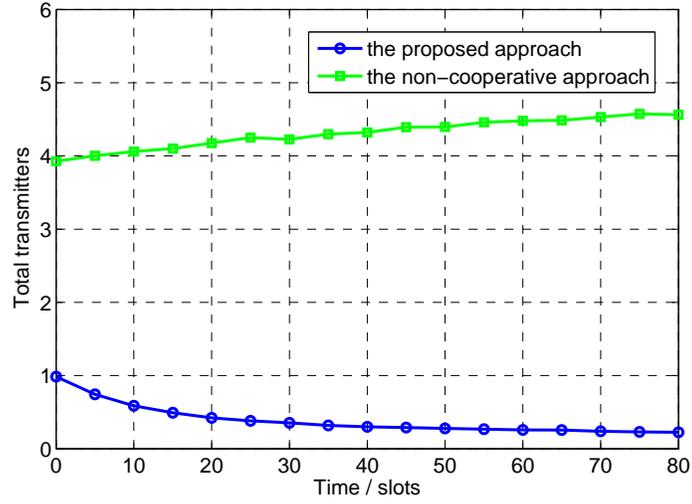}
\caption{Total transmitters by the proposed approach and the non-cooperative approach as a function of time for networks with $N=8, L=800m, D=250m$.} \label{transmitter}
\end{figure}

In Fig.~\ref{transmitter}, we show the number of total transmitters as a function of time by both the proposed approach and the non-cooperative approach for networks with $N=8, L=800m,$ and $D=250m$. As the vehicles deviate from each other and split into disconnected groups with time, the proposed approach will limit the inefficient broadcastings of OBUs on the edge. However, in the non-cooperative approach, the broadcasting vehicles may even increase with time as the spectrum becomes less crowed when they vehicles split into groups. Thus, the transmitters in the proposed approach decreases with time, while the transmitters in the non-cooperative approach increases.

\begin{figure}[!t]
\centering
\includegraphics[width=4.2in]{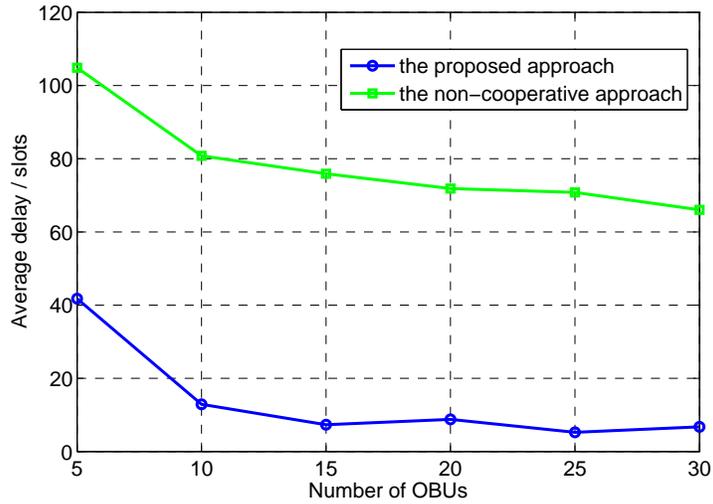}
\caption{Average delay by the proposed approach and the non-cooperative approach as a function of the number of OBUs $N$ for networks with $L=100 \times N (m), D=250m$.} \label{average_delay}
\end{figure}

In Fig.~\ref{average_delay}, we show the average delay as a function of the number of OBUs $N$ by both the proposed approach and the non-cooperative approach for networks with $L=100 \times N(m),$ and $D=250m$. We can see that the average delay decreases as the number of OBUs increases, and our proposed algorithm achieves much better performances than the non-cooperative approach. With more OBUs included in the network, the packets can be transmitted or received by more vehicles, and the performance in average delay is thus improved. Further, the increasing number of OBU also lowers the probability of splitting, which decreases the average delay in another way. As the proposed approach aims at minimizing the average delay formulated in the utility function given by (\ref{utilityS}), the proposed approach again has a better performance.

\begin{figure}[!t]
\centering
\includegraphics[width=4.2in]{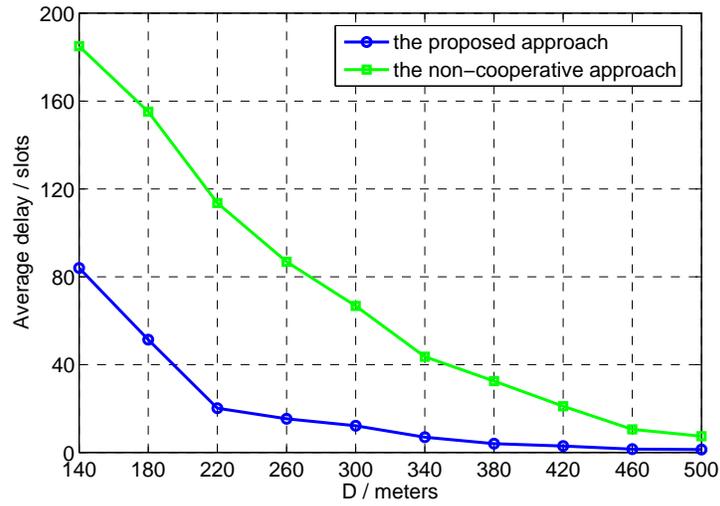}
\caption{Average delay by the proposed approach and the non-cooperative approach as a function of the diameter of RSU coverage $D$ for networks with $N=8, L=800m$.} \label{average_delay2}
\end{figure}

In Fig.~\ref{average_delay2}, we show the average delay as a function of the diameter of the RSU coverage $D$ by the proposed approach and the non-cooperative approach for networks with $N=8$, and $L=800m$. As the diameter of the RSU $D$ is enlarged, or equivalently the V2R transmission time is increased, the initial possessed packets are increased for each OBU in the V2V phase, which directly reduces the time for P2P transmissions, as shown in Fig.~\ref{average_delay2}. When the coverage of the RSU is extremely large, or equivalently the V2R transmission time is quite sufficient, the OBUs may achieve the entire popular file in the V2R phase, which explains why both the two curves converge to $0$ when $D$ surpasses a certain threshold. In this situation, the PCD problem degenerates into pure broadcasting services.

\begin{figure}[!t]
\centering
\includegraphics[width=4.2in]{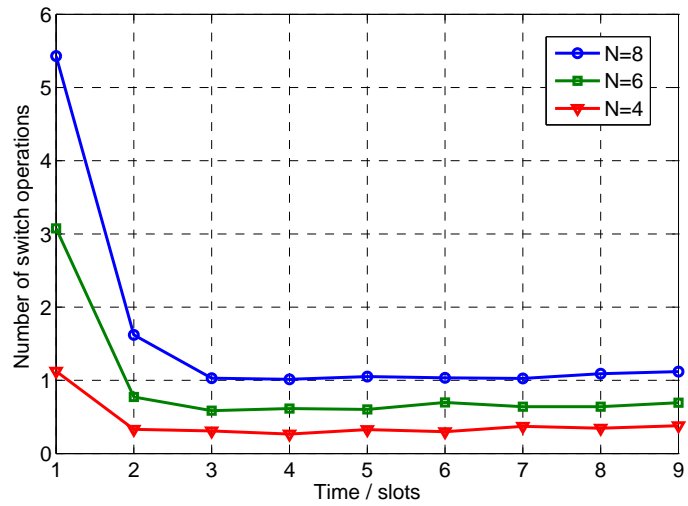}
\caption{Number of switch operations in the proposed approach as a function of time for networks with $D=250m, N=4,6,8, L=100 \times N (m)$}\label{convergence}
\end{figure}

In Fig.~\ref{convergence}, we show the number of switch operations in the proposed scheme as a function of time for networks with $N=4,6,8$. The complexity of our proposed algorithm is mainly caused by the switch operations, the number of which decreases rapidly in the first few slots and then stays at a very low level, as shown in Fig.~\ref{convergence}. The small number of switch operations implies that the proposed approach can adapt to the environmental changes only with limited complexity. Also, we find that the stable complexity, represented by the average number of switch operations in table slots, increases linearly with the number of OBUs $N$, which may imply that the network scale can be much larger in practical systems.

%%%%%%%%%%%%%%%%%%%%%%
\section{Conclusions}%
%%%%%%%%%%%%%%%%%%%%%%

In this paper, we address the PCD problem in vehicular ad hoc networks, where the OBUs may not finish downloading a large file directly from the RSU when they are moving at high speeds. For completing the file delivery process, a P2P network is constructed in the V2V phase of PCD, for which we propose a cooperative approach based on the coalition formation games. In the proposed approach, we formulate the performance of the average delay into a utility function, which highly decides the individual profits for each OBU in the proposed coalition formation game. With the proposed coalition formation algorithm, the OBUs dynamically converge to a Nash-stable partition, and the most efficient coalition in the final partition is allowed to broadcast in the current slot. In the simulation part, we further propose a non-cooperative approach in which the OBUs broadcast as long as the spectrum is unoccupied. We show that the proposed protocol achieves much better performances in the average delay and power efficiency, compared with the non-cooperative approach.

%\renewcommand{\baselinestretch}{1.2}

%%%%%%%%%%%%%%%%%%%%%%%%%%%%%%%%
%%table
%%%%%%%%%%%%%%%%%%%%%%%%%%%%%%%%

\begin{table}[!t]
\renewcommand{\arraystretch}{2.0}
\caption{The Coalition Formation Algorithm for On-board Units (OBUs) in the V2V phase} \label{coalitionFormationAlgorithm} \centering
\begin{tabular}{p{150mm}}

\hline

Given any partition $\Pi_{initial}$ of the OBUs set $\Omega$ with the initialized history collections \\
$H(i) = \emptyset, \forall i \in \Omega$, the OBUs engage in coalition formation algorithm as follows: \\
$\ast$ \textbf{repeat} \\
\quad \quad For a randomly chosen OBU $i \in \Omega$, with current partition $\Pi_{current}$ ($\Pi_{current}$ \\
\quad \quad $= \Pi_{initial}$ in the first round) \\
\quad \quad \quad a) Search for a possible switch operation from $S_{\Pi}(i) = S_m, m \in \{1,2,\ldots,l\}$ \\
\quad \quad \quad to $S_k \in {\Pi} \cup \{\emptyset\}, S_k \ne S_{\Pi}(i)$, where $S_k \cup \{i\} {\succ}_i S_{\Pi}(i)$ and $S_k \cup \{i\} \notin H(i)$. \\

\quad \quad \quad b) If such switch operation exists, OBU $i$ performs the follows steps: \\
\quad \quad \quad \quad b.1) Update the history collection $H(i)$ by adding coalition $S_{\Pi_{current}}(i)$, \\
\quad \quad \quad \quad before leaving it. \\
\quad \quad \quad \quad b.2) Leave the current coalition $S_{\Pi_{current}}(i)$.\\
\quad \quad \quad \quad b.3) Join the new coalition $S_{\Pi_{next}}(i)$ that improves its payoff.\\
$\ast$ \textbf{until} the partition converges to a final Nash-stable partition $\Pi_{final}$. \\

\hline

\end{tabular}
\end{table}

\begin{table}[!t]
\caption{The Splitting Scheme for the On-board Units (OBUs) in the Network}
\label{split}
\renewcommand{\arraystretch}{2.0}
\centering
\begin{tabular}{p{150mm}}

\hline

For any OBU $i \in \Omega$, it performs the following steps: \\
\quad a) Discover its neighbors, where the well-known neighboring discovery algorithms \\
\quad in ad hoc routing discovery \cite{MM-2004} can be engaged.\\
\quad b) Discover the nearby subnetworks that have at least one member as its neighbor \\
\quad by asking from its neighbors \\
\quad c) Join the largest nearby subnetwork with the number of members below $N_{max}$. \\
\quad d) If there is no subnetworks, or all the subnetworks have the maximal members, \\
\quad OBU $i$ establishes a new subnetwork with itself as the only member. \\

\hline

\end{tabular}
\end{table}

\begin{table}[!t]
\renewcommand{\arraystretch}{2.0}
\caption{The Protocol for Popular Content Distribution in Vehicular Ad Hoc Networks} \label{contentDistributionProtocol} \centering
\begin{tabular}{p{150mm}}
\hline

\textbf{Period \uppercase\expandafter{\romannumeral1}: V2R Phase}  \\
\quad The RSU periodically broadcasts the packets of popular file $\Gamma = \{\gamma_0,\ldots,\gamma_{M-1}\}$, \\
\quad and the OBUs in $\Omega$ keep receiving data from the RSU when they pass through.\\

\textbf{Period \uppercase\expandafter{\romannumeral2}: V2V Phase}  \\

\quad With the initially possessed packets given by (\ref{initialSegmentNumber}) and (\ref{firstReceivedIndex}), the OBUs periodically\\ \quad perform the following three stages: \\

\quad \emph{Stage \uppercase\expandafter{\romannumeral1}: Network Splitting}  \\
\quad \quad The OBUs perform the splitting scheme in Table \ref{split} every $K$ slots, and the network \\
\quad \quad may split into smaller subnetworks.\\
\quad \emph{Stage \uppercase\expandafter{\romannumeral2}: Coalition Formation}  \\
\quad \quad For each subnetwork, the OBUs perform the coalition formation algorithm in Table \\
\quad \quad \ref{coalitionFormationAlgorithm} to obtain the final partition $\Pi_{final}$ in the current slot. \\
\quad \emph{Stage \uppercase\expandafter{\romannumeral3}: Data Broadcasting}  \\
\quad \quad For each subnetwork, the OBUs in coalition $S$ broadcast the packets resulting \\
\quad \quad from the greedy algorithm proposed in Section \uppercase\expandafter{\romannumeral3}, where $S \in \Pi_{final}$ is the coalition \\
\quad \quad with the highest service rate given by (\ref{serviceRateExpression}). \\

\hline
\end{tabular}
\end{table}

\begin{table*}
\begin{center}
\caption{Parameters for Simulation} \label{parameters}
\begin{tabular}{|c|c|}

\hline $T = 100ms$ & the periodicity of the V2V phase\\
\hline $N = 5 \sim 30$ & the number of OBUs in the network\\
\hline $L = 500m \sim 3000m$ & the length of the vehicle fleet\\
\hline $N_{max} = 8$ & the maximal members in a subnetwork\\
\hline $K = 10$ & the periodicity of the splitting scheme\\
\hline $D = 140 \sim 500m$ & the diameter of the RSU's coverage\\
\hline $\alpha = 100, \beta = 1$ & the pricing factors \\
\hline $M = 100$ & the number of entire packets\\
\hline $Ms = 100Mb$ & the size of the popular file\\
\hline $v_{min} = 20 m/s$ & the minimal speed \\
\hline $v_{max} = 40 m/s$ & the maximal speed \\
\hline $d_{min} = 100m$ & the security distance \\
\hline $d_{max} = 1000m$ & the maximal distance \\
\hline $a = 1 m/s^2$ & the acceleration \\
\hline $p = 0.1$ & the probability of changing speed \\
\hline $W = 30MHz$ & the channel bandwidth\\
\hline $c_0 = 5Mb/s$ & the V2R channel rate\\
\hline $\eta = 10^6$ & the signal-to-noise rate at the transmitter\\
\hline $\kappa = 10dB$ & the power ratio of LOS against non-LOS\\

\hline
\end{tabular}
\end{center}
\end{table*}

%%%%%%%%%%%%%%%%%%%%%%%%%%%%%%%%%%%%%%%%%%%%%%%%%
%% figure
%%%%%%%%%%%%%%%%%%%%%%%%%%%%%%%%%%%%%%%%%%%%%%%%%

\end{document}